\documentclass{IEEEtran}    

\usepackage{hyperref,graphicx}
\usepackage{longtable}

\usepackage{longtable}
\usepackage{latexsym}
\usepackage{amsmath}
\usepackage{amscd}
\usepackage{amsfonts}
\usepackage{color}
\usepackage{float}
\usepackage{longtable}
\usepackage{graphicx}
\usepackage{xspace}
\usepackage{stmaryrd}
\usepackage[utf8]{inputenc}
\usepackage{color}

\DeclareMathAlphabet{\mathsl}{OT1}{cmr}{m}{sl}
\DeclareMathAlphabet{\mathsc}{OT1}{cmr}{m}{sc}
\DeclareMathAlphabet{\mathslbf}{OT1}{cmr}{bx}{sl}
\DeclareFontFamily{OT1}{pzc}{}
\DeclareFontShape{OT1}{pzc}{m}{it}%
             {<-> s * [1.150] pzcmi7t}{}
\DeclareMathAlphabet{\mathscript}{OT1}{pzc}{m}{it}

\newtheorem{thm}{Theorem}[section]
\newtheorem{lem}[thm]{Lemma}
\newtheorem{cor}[thm]{Corollary}
\newtheorem{propo}[thm]{Proposition}
\newtheorem{clm}[thm]{Claim}
\newtheorem{defn}[thm]{Definition}
\newtheorem{assm}[thm]{Assumption}
\newtheorem{rem}[thm]{Remark}
\newtheorem{obs}[thm]{Observation}
\newtheorem{egs}[thm]{Example}

\newtheorem{fct}[thm]{Fact}
\newtheorem{cons}[thm]{Construction}
\newtheorem{nte}[thm]{Note}

\newenvironment{lemma}{\begin{lem}}{\end{lem}}

\newcommand{\figref}[1]{Figure~\ref{#1}}

\renewcommand{\eqref}[1]{\mbox{Equation~(\ref{#1})}}

\def\figurewidth{0.97\columnwidth}

\newcommand{\widthfigure}[3]{\begin{figure}\begin{center}\begin{tabular}{|p{\figurewidth}|}\hline#1\hline\end{tabular}\end{center}\smallskip\caption{#2.}\label{#3}\end{figure}}

\newcommand{\authnote}[2]{
\ifnum\authnotes=1 
  \begin{center}
    \fbox{\begin{minipage}{5.7in}
      \textbf{#1 says:} #2
    \end{minipage}}
  \end{center} 
\fi
}

\newcommand{\veca}{\mathbf{a}}

\newcommand{\vecx}{\mathbf{x}}

\newcommand{\Zq}{\mathbb{Z}_q}

\def\getsr{\stackrel{{\scriptscriptstyle\$}}{\leftarrow}}

\newcommand{\pmul}{\ensuremath{\pi_{\mathsf{DM}}}\xspace}

\newcommand{\pmmul}{\ensuremath{\pi_{\mathsf{DMM}}}\xspace}
\newcommand{\pip}{\ensuremath{\pi_{\mathsf{IP}}}\xspace}

\newcommand{\pcomp}{\ensuremath{\pi_{\mathsf{DC}}}\xspace}
\newcommand{\pdecomp}{\ensuremath{\pi_{\mathsf{Decomp}}}\xspace}

\newcommand{\fti}[1]{\ensuremath{\mathcal{F}^{\mathcal{D}_{#1}}_{TI}}\xspace}

\newcommand{\shareq}[1]{\ensuremath{\llbracket{#1}\rrbracket_{_q}}\xspace}
\newcommand{\sharetwo}[1]{\ensuremath{\llbracket{#1}\rrbracket_{_2}}\xspace}

\begin{document}

\title{VirtualIdentity: Privacy-Preserving User Profiling
\thanks{A preliminary version of this work appeared at the 2016 IEEE/ACM International Conference on Advances in Social Networks Analysis and Mining (ASONAM).}
}
\author{Sisi~Wang, Wing-Sea~Poon,  Golnoosh~Farnadi, Caleb~Horst,  Kebra~Thompson,\\ Michael~Nickels, Rafael~Dowsley, Anderson~C.~A.~Nascimento and  Martine~De~Cock

\thanks{Sisi Wang, Wing-Sea Poon, Caleb Horst, Kebra Thompson, Michael Nickels, Anderson~C.~A.~Nascimento and Martine De Cock are with the Institute of Technology, University of Washington Tacoma. E-mails: \{sisiwang, wpoon93, calebjh, kebrat, mnickels, andclay, mdecock\}@uw.edu.}
\thanks{Golnoosh Farnadi is with the Dept.~of Computer Science, University of California at Santa Cruz. Email: gfarnadi@ucsc.edu.}
\thanks{Rafael~Dowsley is with the Department of Computer Science, Aarhus University. Email: rafael@cs.au.dk. Rafael Dowsley has received funding from the European Research Council (ERC) under the European  Union's  Horizon  2020  research  and  innovation programme  under  grant  agreement  No  669255  (MPCPRO).}
\thanks{Martine De Cock is with the Dept.~of Applied Mathematics, Computer Science and Statistics, Ghent University. Email: mdecock@ugent.be.}
}

\maketitle

\begin{abstract}
User profiling from user generated content (UGC) is a common practice that supports the business models of many social media companies. Existing systems require that the UGC is fully exposed to the module that constructs the user profiles. In this paper we show that it is possible to build user profiles without ever accessing the user's original data, and without exposing the trained machine learning models for user profiling -- which are the intellectual property of the company -- to the users of the social media site. We present VirtualIdentity, an application that
uses secure multi-party cryptographic protocols to detect the age, gender and personality traits of users by classifying their user-generated text and personal pictures with trained support vector machine models in a privacy-preserving manner.
\end{abstract}


\IEEEpeerreviewmaketitle

\section{Introduction}
As more users are creating their own content on the web, there is a growing interest to mine this data for use in personalized information access services, recommender systems, tailored advertisements, and other applications that can benefit from personalization \cite{KOSINSKI:2013}. 
In addition to myriad applications in e-commerce, there is a growing interest in user profiling for digital text forensics \cite{pan}. Furthermore, the popularity of applications such as How-Old.net and HowHot.io shows that users are directly interested in their own personal features analysis as well \cite{microsoft,howhot}. What is common across all of these existing personalized services is that the personal data of users, such as their pictures and text, is fully exposed to the user profiling service. 

\begin{figure}
\centering
\includegraphics[width=3.2in]{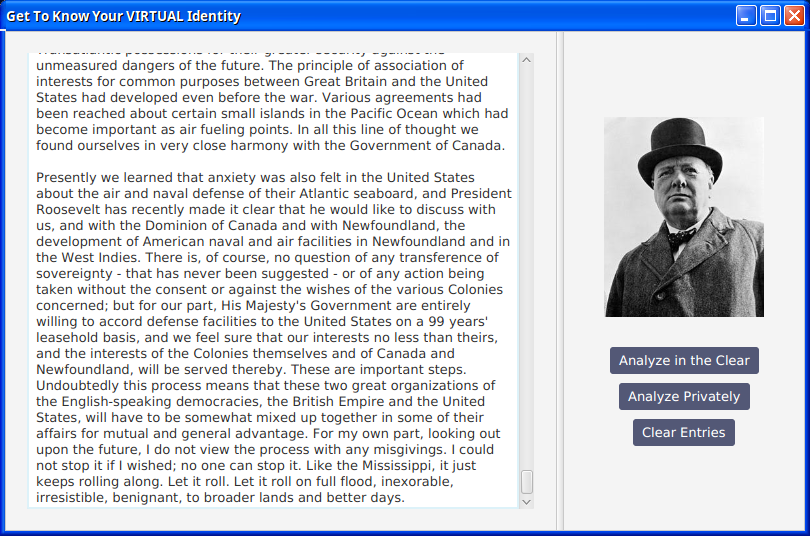}

\footnotesize{(a) The user inputs text and a profile picture.}

\vskip 10pt

\includegraphics[width=3.2in]{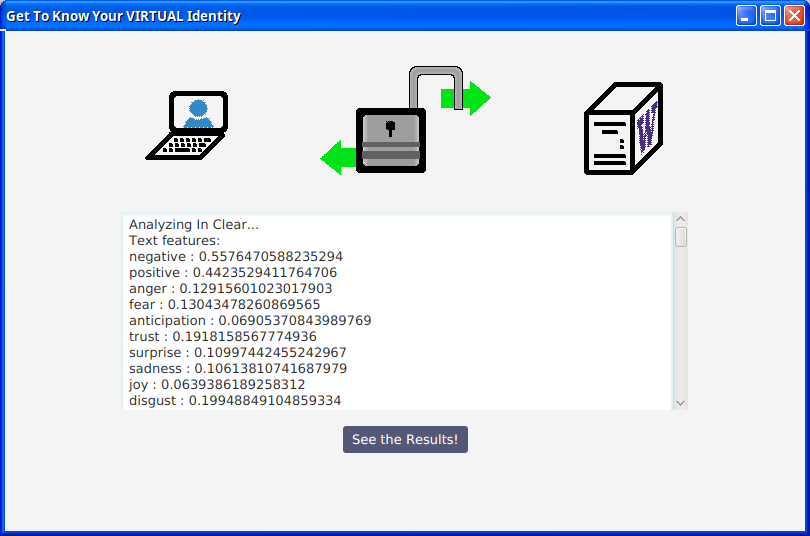}
\includegraphics[width=3.2in]{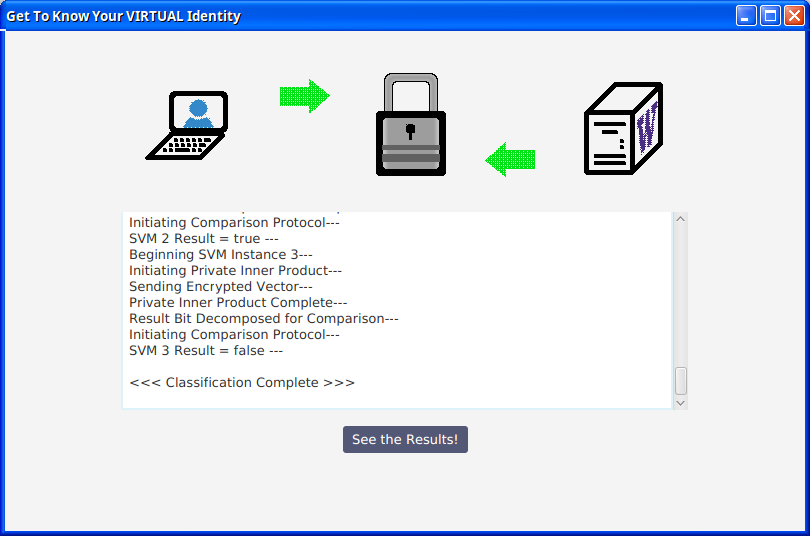}

\footnotesize{(b) For demo purposes, the analysis is done both in the clear\\ and in a privacy-preserving manner.}

\vskip 10pt

\includegraphics[width=3.2in]{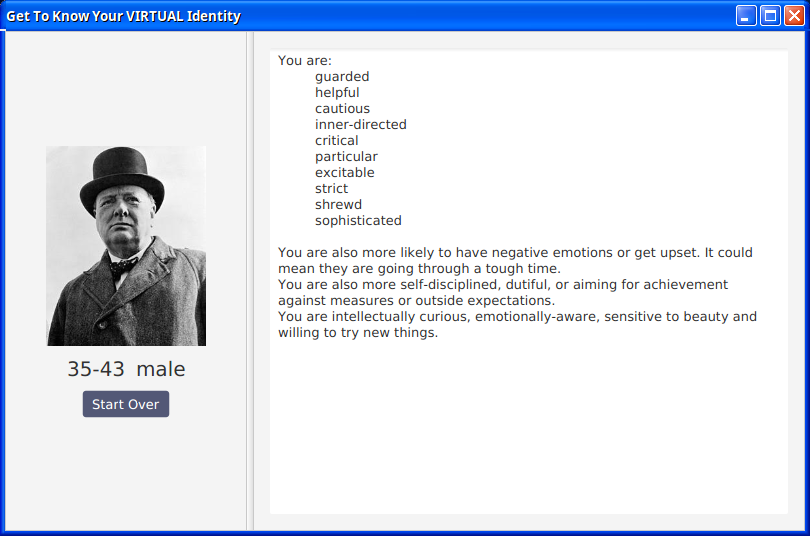}

\footnotesize{(c) The service returns age, gender, and personality analysis.}

\caption{Screenshots of VirtualIdentity application.\label{SCREEN1}}
\end{figure}

An obvious way to circumvent this would be to perform the user profiling entirely on the user's side. However, this would imply sharing proprietary, trained machine learning models for user profiling with each user of the social media site. Applying traditional cryptography to encrypt the personal data of the user (henceforth called the client) before sending it to the user profiling service (the service, or server) is not a solution either, as data encrypted with usual techniques becomes useless, and user characteristics can no longer be derived from it. Hiding the client's data from the service, while still allowing the client to use the service, requires novel cryptographic techniques that not only protect private information but also allow mathematical operations to be performed on encrypted data. To this end, the VirtualIdentity application that we present in this paper (see Figure \ref{SCREEN1}) relies on secure multi-party computation, a process in which client and server jointly compute classification labels by exchanging encrypted messages, while keeping their own inputs private. As a result, VirtualIdentity allows a user to run our trained support vector machines (SVMs) for detection of age, gender, and personality traits, without leaking any personal text or profile picture to our server. In addition, the user does not learn anything about the coefficients of our SVM models.

Other services exist that will predict a user's age, gender, or personality based on UGC. For example, users can input their tweets or text and receive back scores of their personality, needs, and values \cite{ibm}. Another site allows users to input a photo and receive an estimation of the gender and age of each face in the photo \cite{microsoft} while a third estimates the user's attractiveness and age from a photo \cite{howhot}. However, none of these services attempt to keep the user's data private. To the best of our knowledge, VirtualIdentity is the first platform to construct user profiles while preserving both the privacy of the user's data and the prediction models.

This paper is an extended version of the work that appeared at the 2016 IEEE/ACM International Conference on Advances in Social Networks Analysis and Mining (ASONAM) \cite{asonam}. This full version contains the description of the cryptographic tools that are used.

%
%

\section{Predictive Models}
Much work has been done recently using machine learning classification to predict age, gender, and personality based on images and text ``in the clear'', i.e.~without any attempts for privacy preservation. In this paper we use SVMs, which are known as state-of-the-art classification techniques for detecting age, gender and personality traits from text and images \cite{linguistic,unfiltered,automatic,computational,ageest}.
\subsection{Age and Gender Classification}

For age and gender classification we used the IMDB image dataset, which is part of the IMDB-WIKI dataset. This set is formed by 460,723 face images crawled from IMDB websites with age and gender information \cite{imageDataset}. From each image, we detected face and cropped the margin to $40\%$ using OpenCV \cite{opencv}. Then we extracted 136 facial landmark features using Dlib \cite{dlib}. These features, which include attributes such as the exact locations of the eyes, nose, and mouth, were then used to train the models. Some facial images were dropped because there is no face or more than one face detected in them by OpenCV or Dlib's facial landmark detector. In addition, images with unreasonable age (e.g.,negative age) and images without gender information were taken out as well. After preprocessing and feature extraction, we have 318,562 valid instances remaining in the set. The set is divided into 4 similar-sized age groups: (7-26), (27-34), (35-43), (44-101). For age classification, each instance will be classified into one age bucket. For the actual training, we used 6000 of the IMDB dataset images such that the age and gender distributions of the selected images are representative of the full set.

We trained a binary SVM classifier for gender classification, and three binary SVMs for age classification. We use the results of all three age classifiers to determine the most likely age bracket. When a new instance comes in, it will first be classified by Age\underline{{ }{ }}SVM2 to determine if it is younger than 35 or not. If it is classified into younger than 35-year-old group, it will then be scored against Age\underline{{ }{ }}SVM1 to see if it is younger than 27. Otherwise, Age\underline{{ }{ }}SVM3 will be applied to check if the new instance is older than 43. This approach is similar to the approach of Han et al. \cite{ageest}. While they use additional models to then predict an actual age inside the bracket, we return the result determined from the three original SVMs. 

Table 1 depicts the baseline values compared with the average accuracies of the gender SVM and three age SVMs using 10-fold cross-validation.
\begin{table*}[!ht]
\centering
\includegraphics[width=4.5in]{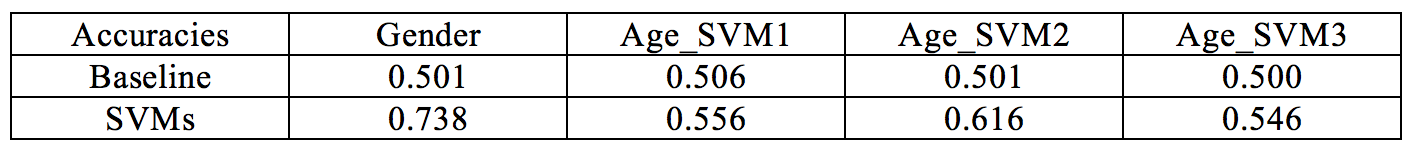}
\caption{Age and Gender Classification SVMs Accuracy Table\label{ACCURACYTABLE1}}
\end{table*}

\subsection{Personality Traits Detection}

For personality we report scores using the traits of the widely accepted model, the Big Five, consisting of the following five results:
openness to experience, conscientiousness, extraversion, agreeableness, and neuroticism \cite{big5}. Our goal is to create classifiers to predict if one user displays those 5 characteristics. The dataset we used for personality traits detection is a dataset with 2467 essays (one empty instance was removed from the original 2468) from psychology students who were told to write whatever came to their mind for 20 minutes \cite{linguistic}. Each essay was analysed and given Big Five personality ground truth labels by Pennebaker et al.~\cite{liwc1999}. 


We extracted three kinds of features from the essays as input for the classifiers: 14 MRC features, 10 NRC features, and 19 LIWC features (43 features in total). 

MRC is a psycholinguistic database which contains psychological and distributional information about words such as the number of letters in the word, the concreteness, and the age of acquisition~\cite{mrc}.We used the same 14 MRC features as Farnadi et al. \cite{computational} The features are: number of letters in the word (NLET), number of phonemes in the word (NPHON), number of syllables in the word (NSYL), Kucera and Francis written frequency (KF FREQ), Kucera and Francis number of categories (KF NCATS), Kucera and Francis number of samples (KF NSAMP), Thorndike-Lorge frquency (TL FREQ), Brown verbal frequency (BROWN FREQ), Familiarity (FAM), concreteness (CONC), imagery (IMAG), mean Colerado Meaningfulness (MEANC), mean Pavio Meaningfulness (MEANP), and age of acquisition (AOA). Each feature is computed by averaging the feature value of all the words in the essay.~\cite{computational}

%
NRC is a lexicon that contains more than 14,000 distinct English words annotated with 8 emotions (anger, fear, anticipation, trust, surprise, sadness, joy, and disgust), and 2 sentiments (negative, positive)\cite{semantic}. For each document we counted the number of words in each of the 8 emotion and 2 sentiment categories, resulting in 10 features per document. 

The Linguistic Inquiry and Word Count tool (LIWC) is a well-known text analysis software which is widely used in psychology studies \cite{liwc}. Part of the LIWC features rely on a proprietary dictionary. Our SVM models are trained on 19 LIWC features that relate to standard counts and that do not require the specific LIWC dictionary: word count, words per sentence, number of unique words, number of words longer than six letters, number of abbreviations, emoticons, question marks, periods, commas, colons, semi-colons, exclamation marks, dashes, quotation marks, apostrophes, parentheses, other punctuation marks, all punctuation marks, and number of interrogative sentences.

Since more than one trait can be present in the same user, we used the 43 features to trained one binary SVM classifier for each of the five traits, which separates the users displaying the characteristic from those who do not. Table 2 depicts the baseline values compared with the average accuracies of our 5 SVMs using 10-fold cross-validation.
\begin{table*}[!ht]
\centering
\includegraphics[width=4.5in]{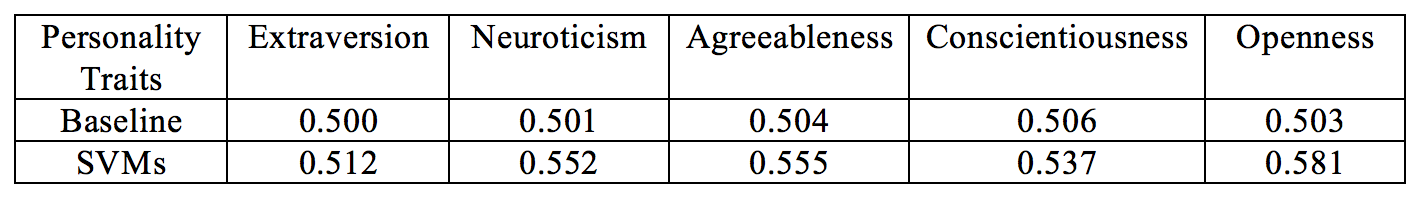}
\caption{Personality Classification SVMs Accuracy Table\label{ACCURACYTABLE2}}
\end{table*}

%

All SVMs in our model bank were trained using scikit-learn in Python, with a linear kernel and penalty parameter $C=1$.

\begin{figure*}
\centering
\includegraphics[width=2\columnwidth]{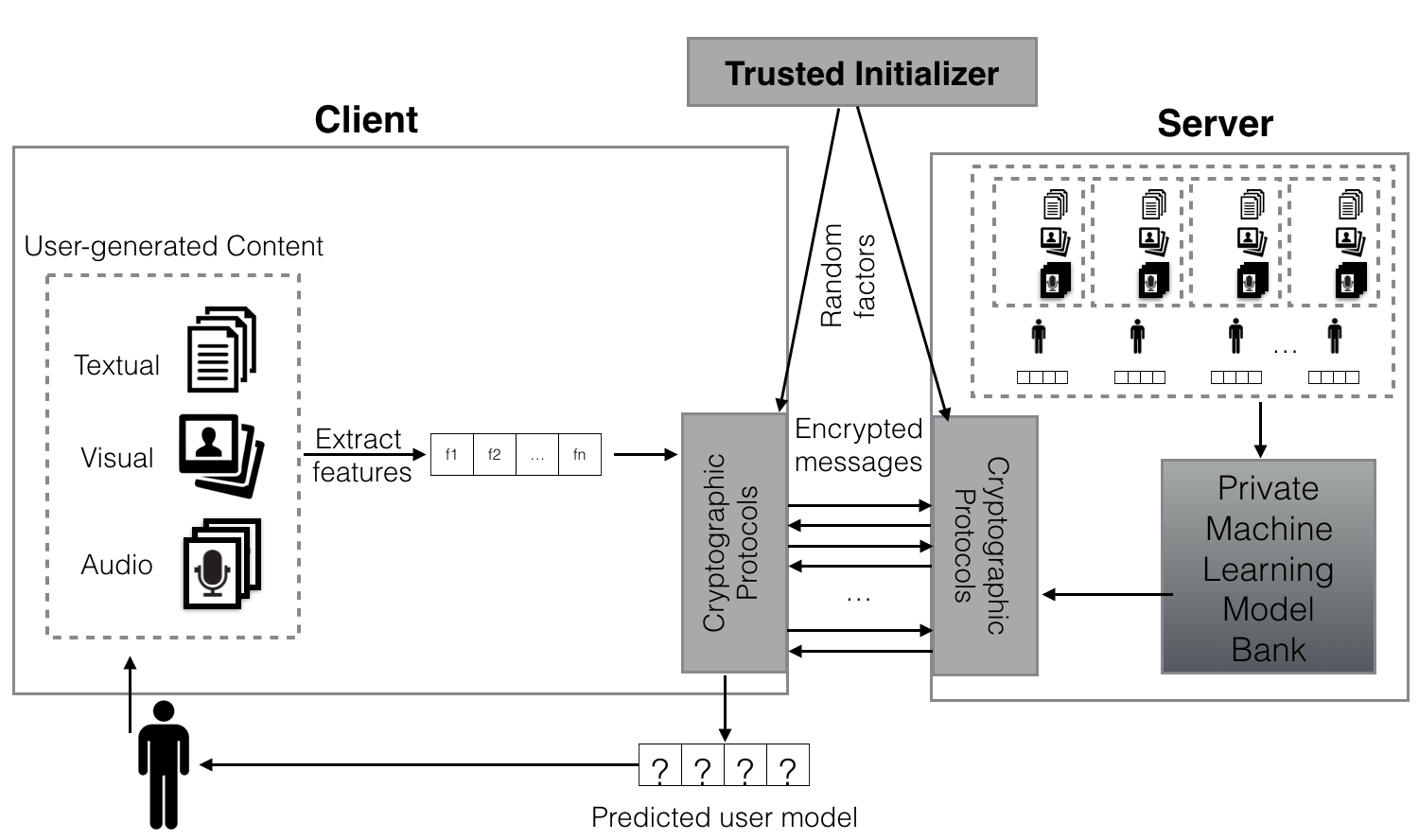}
\caption{System overview of the VirtualIdentity application.\label{OVERVIEW}}
\end{figure*}

The trained SVMs are part of a private machine learning model bank that resides on the server, as shown on the right side in Figure \ref{OVERVIEW}. When a user requests analysis of a snippet of text and a picture, the features described above are extracted from the text and the image on the client side, as shown on the left side in Figure \ref{OVERVIEW}. Neither the user's text, nor the user's image, nor any of the extracted features are leaked to the server. Instead, both the client and the server engage in cryptographic protocols and exchange encrypted messages that ultimately allow the server to classify the feature vectors of the client, without ever seeing them in the clear, as we explain in Section \ref{SEC:SECURE}.

%
%

\section{Adding Privacy to our Classifiers}\label{SEC:SECURE}

Only a limited amount of work has been done in cryptographically secure privacy-preserving machine learning classification and none of it is aimed specifically at user profiling. 

Cryptographically secure privacy-preserving SVM classification protocols have been proposed in \cite{secureSVM,BPTG15,DDKN15,CDHKNNP16}.
The basic idea behind these protocols is to decompose the task of scoring an SVM into smaller tasks and to implement each one of them in a privacy-preserving way. To better understand these previous approaches we recall that 
for the case of two classes, the process for \emph{SVM classification in the clear} is as follows \cite{CORTES:1995}: the client holds an $n$-dimensional input feature vector $\vecx$, and the server holds a trained model that consists of a $n$-dimensional vector of weights $\veca$ and a real number $b$ learned from the training data. The result of the classification is obtained by computing $$\mathsf{sign}\left( \langle \vecx, \veca \rangle -b\right),$$ where the function $\mathsf{sign}\left(y \right)$ gives $+$ if $y > 0$ and $-$ otherwise. For instance, in the case of personality prediction, $\vecx$ is a 43-dimensional vector with features extracted from the client's text and $(\veca, b)$ are the weights and the bias that make up the trained SVM model for e.g.~``neuroticism''. A classification outcome $+$ means that the user is neurotic, and an outcome of $-$ means that he is not. Therefore, to score SVMs privately, one needs to build privacy-preserving for two tasks: computing inner products and performing comparisons.

In \cite{secureSVM}, private inner products and comparisons are obtained by using additive homomorphic encryption and oblivious transfer, while in \cite{BPTG15} the proposed protocols are based on Paillier encryption - a specific kind of additive homomorphic encryption scheme. These operations are usually expensive from a computational complexity point of view, demanding costly modular exponentiations. In \cite{DDKN15,CDHKNNP16}, highly efficient protocols for privacy-preserving comparison and argmax in the commodity-based model \cite{Beaver97} were proposed. In the commodity-based model correlated data is pre-distributed to the parties by a trusted initializer during an off-line setup phase. Here we use the comparison protocol of \cite{CDHKNNP16} combined with a generalization for matrices of Beaver's multiplication protocol \cite{C:Beaver91b} in order to obtain the first implementation of a practical system for solving the problem of privacy-preserving user profiling. 

We have already mentioned how we perform the private classification of personality traits. Now, we briefly describe how we proceed to obtain age and gender prediction. For age prediction, we first split the age groups into $4$ classes, such that the frequency of each class is equal. Because there are $4$ classes, there will be $3$ splitting points. Our target functionality is the one that first determine if the instance is in the lower or upper half by using the binary SVM that is the middle, and then use one binary SVM more in the remaining half to determine the exact class. In the protocol, instead of running the SVMs sequentially, we run all the three independent SVMs in parallel up to the point right before the opening of the results. We then open the result of the SVM in the middle, and after that we open the result of the SVM that is relevant for the chosen half in order to determine the class. A separate SVM is evaluated in a privacy-preserving way to determine the gender of the user. It should be noted, that the techniques used here for implementing privacy-preserving inner product and comparison protocols only work for integer values. To account for this, real values must be converted into integers and lose some of the precision allowable by floating notation.

In this work we consider honest-but-curious adversaries (i.e., adversaries that follow the protocol instructions but try to learn additional information), as done in the other works on privacy-preserving classification.

\subsection{Computing with Secret Sharing}\label{secss}

We use the paradigm of secure computation based on secret sharings. For each shared value, Alice and Bob hold uniformly random values (i.e., the shares) constrained to the condition that they sum up to the actual value that is shared. The computation is then done over the shares, and when it is finished Alice and Bob exchange their shares of the output in order to recover it. In more details, if the shares are in a ring $\Zq$ and the shared value is $x \in \Zq$, Alice and Bob hold uniformly random $x_A, x_B \in \Zq$, respectively, subject to the constraint that $x_A + x_B =x \mod{q}$. Our notation for secret sharings will follow the one used in \cite{CDHKNNP16}.
Let $\shareq{x}$ denote the secret sharing $(x_A, x_B)$. Given secret sharings $\shareq{x}$, $\shareq{y}$, performing additions and subtractions of the shared values, and adding a/multiplying by a constant are very simple operations that can be performed locally by Alice and Bob by just performing the respective operations in their local shares. These local operations will be denoted by $\shareq{z} \gets \shareq{x}+\shareq{y}$, $\shareq{z} \gets \shareq{x}-\shareq{y}$, $\shareq{z} \gets c\shareq{x}$ and $\shareq{z} \gets \shareq{x}+c$. The notation is extend straightforwardly to element-wise secret sharing of vectors $\shareq{\vecx}$ and matrices $\shareq{X}$; and similarly for the operations.

\subsection{Commodity-based Cryptography}\label{seccomm}

Our solution works in the commodity-based model~\cite{Beaver97,C:Beaver95} -- a setup assumption in which a trusted initializer pre-distributes correlated randomness to Alice and Bob during an initial setup phase. This pre-distributed data is independent from the protocol inputs, which can even be fixed far after. We should also emphasize that the trusted initializer does not participate anymore after given the pre-distributed correlated randomness to Alice and Bob. The commodity-based model allows the design of practical, unconditionally secure protocols for many interesting functionalities, for example: inner product~\cite{WISA:DGMN10},  linear algebra~\cite{DDvGMNP15}, oblivious transfer~\cite{Beaver97}, oblivious polynomial evaluation~\cite{IJIS:TNDMIHO15}, verifiable secret sharing~\cite{IEICE:DMOHIN11}, set intersection~\cite{TCC:IKMOP13} and string equality~\cite{TCC:IKMOP13}. 
Given its usefulness it was also already used for obtaining privacy-preserving machine learning protocols \cite{DDKN15,AISec:CDNN15,CDHKNNP16}.
In this work the trusted initializer is modeled by an ideal functionality $\fti{}$, which is parametrized by an algorithm $\mathcal{D}$ that samples the correlated data to be pre-distributed to Alice and Bob. See \figref{fig-ti} for details. 

\widthfigure{
\begin{center}
\textbf{Functionality} $\fti{}$
\end{center}

$\fti{}$ runs with Alice and Bob and is parametrized by an algorithm $\mathcal{D}$.
Upon initialization run $(D_A, D_B) \getsr \mathcal{D}$. Deliver $D_A$ to Alice and 
$D_B$ to Bob.\\
\\
}{The Trusted Initializer functionality}{fig-ti}

Another advantage of the commodity-based model is that it is one of the setup assumptions allowing to obtain  UC-security \cite{FOCS:Canetti01}, which is the notion of security allowing the modular design of protocols while keeping the security guarantees (as done in our privacy-preserving protocols). 
\footnote{It is impossible to obtain non-trivial UC-secure two-party computation without setup assumptions \cite{C:CanFis01,STOC:CLOS02}. Some other setup assumptions are also known to allow non-trivial two-party computation, such as: the existence of a common reference string \cite{C:CanFis01,STOC:CLOS02,C:PeiVaiWat08}, of noisy channels \cite{SBSEG:DMN08,JIT:DGMN13}, of a public public-key infrastructure \cite{FOCS:BCNP04}, of signature cards \cite{HofMulUhr05} or of tamper-proof hardware \cite{EC:Katz07,TCC:DotKraMul11,ICITS:DowMulNil15}.}

\subsection{Secure Distributed Matrix Multiplication}\label{secmult}

While Section \ref{secss} described many operations that can be performed locally by Alice and Bob, the most important operation that is missing and that requires interaction between them is the multiplication of shared values. 
This can be an expensive operation in general, but in the commodity-based model there is a very elegant and simple solution by Beaver \cite{C:Beaver91b}. As we will also need secure (distributed) inner product as a building block, we describe here a generalization of Beaver's idea that performs secure distributed matrix multiplication (and so covers both cases of interest) following the description used in \cite{CDHKNNP16}. Alice and Bob hold secret sharings $\shareq{X}$ and $\shareq{Y}$ of matrices $X \in \mathbb{Z}_q^{i \times j}$ and $Y \in \mathbb{Z}_q^{j \times k}$ and they want to obtain a secret sharing corresponding to the matrices' product. The approach is to have the trusted initializer pre-distributing a random matrix multiplication triple to Alice and Bob, i.e., secret sharings $\shareq{U}, \shareq{V}$ and $\shareq{W}$ with $U$ and $V$ uniformly random in $\mathbb{Z}_q^{i \times j}$ and $\mathbb{Z}_q^{j \times k}$, respectively, and $W=UV$. This matrix multiplication triple is then easily derandomize by Alice and Bob in order to match the actual inputs. The protocol $\pmmul$ is described in \figref{prot:distmult}. The protocol correctness can be easily checked using the fact that $W=UV$. The protocol security essentially comes from the fact that in the revealed values, $D$ and $E$, the inputs $X$ and $Y$ are masked by completely random one-time pads $U$ and $V$, respectively (and the one-time pads are only used once). A more detailed security proof can be found in \cite{CDHKNNP16,Dowsley16}:

\begin{lemma}
The protocol $\pmmul$ UC-realizes the distributed matrix multiplication functionality  
against honest-but-curious adversaries in the commodity-based model.
\end{lemma}

\widthfigure{
\begin{center}
\textbf{Secure Matrix Multiplication Protocol $\pmmul$}
\end{center}
The protocol is parametrized by the size $q$ of the ring and the dimensions $i$, $j$ and $k$ of the matrices, and runs with Alice and Bob. The trusted initializer chooses uniformly random 
$U$ and $V$ in $\mathbb{Z}_q^{i \times j}$ and $\mathbb{Z}_q^{j \times k}$, respectively, computes $W=UV$ and pre-distributes secret sharings $\shareq{U}, \shareq{V}, \shareq{W}$ to Alice and Bob. 
Alice and Bob have inputs $\shareq{X}$, $\shareq{Y}$ and interact as follows:\\
\begin{enumerate}
\item Locally compute $\shareq{D} \gets \shareq{X}-\shareq{U}$ and $\shareq{E} \gets \shareq{Y}-\shareq{V}$, then open $D$ and $E$.
\item Locally compute $\shareq{Z} \gets \shareq{W} + E \shareq{U} + D \shareq{V} +DE$.
\end{enumerate}\\
}{The protocol for secure distributed matrix multiplication  \cite{CDHKNNP16}.}{prot:distmult}

\textbf{Notation:} We denote by $\pmul$ the protocol for the special case of multiplication of single elements. The special case of inner product computation will be denoted as $\pip$.

\subsection{Secure Distributed Comparison}\label{sec:comp}

We also use as a building block the secure distributed comparison protocol of \cite{CDHKNNP16}. Alice and Bob want to compare two $\ell$-bit integers, $x = x_\ell \ldots x_1$ and $y = y_\ell \ldots y_1$. Alice and Bob have as input secret sharings 
$\sharetwo{x_i}$ and $\sharetwo{y_i}$ of each bit of $x$ and $y$. The output of the distributed comparison is $\sharetwo{1}$ if $x \geq y$ and $\sharetwo{0}$ if $x < y$. No additional information should be leaked to Alice or Bob. 
The (basic) comparison protocol $\pcomp$ is described in \figref{prot:seccomp}. The security of $\pcomp$ was proved in \cite{CDHKNNP16}. The intuition is that the only non-local operations are the multiplications, so the security of the distributed comparison protocol follows from the security of the distributed multiplication protocol.

\widthfigure{
\begin{center}
\textbf{Secure Distributed Comparison Protocol $\pcomp$}
\end{center}
Let $\ell$ be the bit length of the integers to be compared. The trusted initializer pre-distributes the correlated randomness necessary for the
execution of all instances of the distributed multiplication protocol. Alice and Bob have as inputs shares $\sharetwo{x_i}$ and $\sharetwo{y_i}$ of each bit of $x$ and $y$. 
The protocol proceeds as follows:\\
\begin{enumerate}
\item For $i = 1,\ldots, \ell$, compute in parallel $\sharetwo{d_i} \gets \sharetwo{y_i}\left(1- \sharetwo{x_i}\right)$
using the multiplication protocol $\pmul$ and locally compute $\sharetwo{e_i} \gets \sharetwo{x_i} + \sharetwo{y_i} +1$.
\item For $i = 1,\ldots, \ell$, compute $\sharetwo{c_i} \gets \sharetwo{d_i} \prod_{j=i+1}^\ell \sharetwo{e_j}$ using the multiplication protocol $\pmul$.
\item Compute $\sharetwo{w} \gets  1 + \sum_{i=1}^\ell \sharetwo{c_i}$ locally. 
\end{enumerate}\\
}{The protocol for secure distributed comparison \cite{CDHKNNP16}}{prot:seccomp}

\begin{lemma}[\cite{CDHKNNP16}]
The distributed comparison protocol $\pcomp$ UC-realizes the distributed comparison functionality 
against honest-but-curious adversaries in the commodity-based model.
\end{lemma}

We use the optimized version of the protocol (described in \cite{CDHKNNP16}), which only has $2+ \log{\ell}$ rounds.

\subsection{Secure Bit-Decomposition}\label{sec:decomp}

For obtaining the privacy-preserving SVMs, the secure inner product $\pip$ and the secure distributed comparison $\pcomp$ protocols need to be integrated; however, the inner product will be used with inputs over large rings, while the comparison protocol works on the binary field. Therefore it is necessary to have a protocol for converting secret sharings in the big field $\shareq{x}$ to bit-wise secret-sharings $\sharetwo{x_i}$ in the binary field (for $x=x_\ell \cdots x_1$). This work uses the same specialized bit-decomposition protocol as in \cite{CDHKNNP16} (which is similar to the one of Laud and Randmets \cite{CCS:LauRan15}). It works for $q=2^\ell$ and the main idea is use a carry computation to obtain the bitwise secret sharings $\sharetwo{x_i}$ starting from the shares of $\shareq{x}$ that Alice and Bob have. The (basic version of the) bit-decomposition protocol $\pdecomp$ is presented in \figref{prot:decomp}. The security of $\pdecomp$ follows intuitively from the fact that the only non-local operations are the distributed multiplications, and these are performed using a UC-secure protocol.

\widthfigure{  \begin{center}
\textbf{Secure Bit-Decomposition Protocol $\pdecomp$}
\end{center}
Let $\ell$ be the bit length of the value $x$ to be re-shared. 
All distributed multiplications using protocol $\pmul$ will be over $\mathbb{Z}_2$ and the required correlated randomness is pre-distributed by the trusted initializer. Alice and Bob have as input $\shareq{x}$ for $q=2^\ell$ and proceed as follows:\\

\begin{enumerate}
\item Let $a$ denote Alice's share of $x$, which corresponds to the bit string $a_\ell\ldots a_1$. Similarly, let $b$ denote Bob's share of $x$, which corresponds to the bit string $b_\ell\ldots b_1$.  Define the secret sharings $\sharetwo{y_i}$ as the pair of shares $(a_i,b_i)$ for $y_i=a_i+b_i \mod{2}$, 
$\sharetwo{a_i}$ as $(a_i,0)$ and $\sharetwo{b_i}$ as $(0,b_i)$.

\item Compute $\sharetwo{c_1} \gets \sharetwo{a_1} \sharetwo{b_1}$ using $\pmul$ and locally set $\sharetwo{x_1} \gets \sharetwo{y_1}$.

\item For $i = 2, \ldots, \ell$:
\begin{enumerate}
\item Compute $\sharetwo{d_i} \gets \sharetwo{a_i} \sharetwo{b_i} + 1$
\item $\sharetwo{e_i} \gets \sharetwo{y_i} \sharetwo{c_{i-1}} + 1$
\item $\sharetwo{c_i} \gets \sharetwo{e_i} \sharetwo{d_i} + 1$
\item $\sharetwo{x_i} \gets \sharetwo{y_i} + \sharetwo{c_{i-1}}$
\end{enumerate}
\item Output $\sharetwo{x_i}$ for $i\in \{1,\ldots,\ell\}$.
\end{enumerate}\\
}{The secure bit-decomposition protocol \cite{CDHKNNP16}}{prot:decomp}

\begin{lemma}[\cite{CDHKNNP16}]
Over any ring $\mathbb{Z}_{2^\ell}$, the bit-decomposition protocol $\pdecomp$ UC-realizes the bit-decomposition functionality in the commodity-based model.
\end{lemma}

\textbf{Optimization:} This work use the round-optimized version of $\pdecomp$, which has $2+\lceil \log{\ell} \rceil$ rounds and uses $2\ell \lceil \log{\ell} \rceil + 3 \ell$  instances of the multiplication protocol $\pmul$ over $\mathbb{Z}_2$. Details available in \cite{CDHKNNP16}.

\subsection{Privacy-Preserving SVMs}

As the mentioned before, for a SVM, given the feature vector $\vecx$ and the trained model $(\veca, b)$ that consists of a vector of weights $\veca$ and a real number $b$, the result of the classification is given by
$$\mathsf{sign}\left( \langle \vecx, \veca \rangle - b\right).$$

The idea for obtaining a \emph{privacy-preserving} SVM protocol is as follows: (1) Alice inputs her feature vector $\vecx$ and Bob inputs the weight vector $\veca$ to the secure inner product protocol $\pip$; (2) the resulting secret sharing in $\mathbb{Z}_q$ is then processed by the bit-decomposition protocol $\pdecomp$ to obtain bitwise secret sharings in $\mathbb{Z}_2$; (3) then the bitwise secret sharings can be used in the comparison protocol $\pcomp$ to check whether it is greater than $b$ or not. The final result is then revealed to Alice. 

The security of this protocol follows straightforwardly from the fact that the building blocks are UC-secure (i.e., they can be arbitrarily composed without the security being compromissed) and the fact that no values are ever opened before the final result. In other words, other than the output, each party only sees shares which appear completely random to them.

%
%

\section{System Overview}
The overall architecture of our demo is shown in Figure \ref{OVERVIEW}. The framework consists of a client Java application, a server, and the cryptographic protocols embedded in client and server. Next, we describe these modules.

\subsection{Client Application}
The user interface of our client application shown in Figure \ref{SCREEN1} is developed with JavaFX. The client application consists of a feature extractor and its respective portion of the cryptographic protocols. It allows users to upload user generated content (i.e.~to input written text and to upload a personal picture). It extracts features from the UGC, executes cryptographic protocols with the server, and interprets and displays the final prediction results from the machine learning models. The interpretation of personality refers to Personality Insights \cite{ibm}. 

\subsection{Server}
The server contains its respective  portion of the cryptographic protocols and the private machine learning model bank. The model bank contains the SVMs which are used for predicting personality traits, age and gender.

\subsection{Cryptographic Protocols}
The cryptographic protocols (privacy-preserving protocols for computing inner products and comparisons), are executed in both the client and server side. The trusted initializer pre-distributes correlated data to the client and the server as specified in the commodity-based model during an off-line phase \cite{DDKN15,Beaver97}. The communication between client and server is implemented using sockets. The whole VirtualIdentity application is programmed with Java under JDK 1.8.


\subsection{System Performance}

We have implemented VirtualIdentity within a server of the University of Washington. The client was hosted in a local computer in the city of Tacoma, outside of the university?s network. In average, the running time (including computing and communication delays) was about 5 seconds for the solution in the clear. The privacy preserving solution was evaluated in about 14s-16s. The privacy preserving solution was about 3 times slower than the solution in the clear.

We believe our solution is practical, particularly in the trusted initializer model, where correlated randomness is distributed to the parties in a setup phase.

\section{Conclusion}
Many data-driven personalized services require that private data of users – such as user generated content, personal preferences, browsing behavior, or medical lab results – is scored with proprietary, trained machine learning models. The current widespread practice expects users to give up their privacy by sending their data in the clear to the server where the machine learning models reside. In this paper we have demonstrated that the use of secure multi-party computation techniques allows the construction of user profiles from user generated content while preserving both the privacy of the user’s data and the prediction models. The overall architecture of the VirtualIdentity application is generic and can be extended to
other applications; this would involve extraction of different features and training new models for the private machine learning model bank. 

\bibliographystyle{IEEEtran}

\end{document}